\documentclass{pazh_eng}
\usepackage{epsf}
\usepackage{pstricks}
\newcommand{\ogle}{\mbox{OGLE-1999-BUL-32 }}

\begin{document}
\sloppypar

\title{An upper limit on the X-ray luminosity of the black hole - microlens
OGLE-1999-BUL-32}

\author{\copyright 2001 Ç. M.G. Revnivtsev\inst{1,2}, R.A. Sunyaev\inst{1,2}}

\institute{Space Research Institute RAN, Moscow, Russia
 \and
 Max-Planck-Institut fuer Astrophysik, Garching, Germany}

\authorrunning{REVNIVTSEV, SUNYAEV}
\titlerunning{Upper limits on the X-ray luminosity of OGLE-1999-BUL-32}
\date{ accepted Aug. 31, 2001}
\offprints{revnivtsev@hea.iki.rssi.ru}
\abstract{
We present an upper limit on the 3--20 keV X-ray flux from the black hole -
microlens \ogle, based on RXTE/PCA scans over the Galactic Center region in 
1999-2000. It is shown that the X-ray luminosity of the black hole did not
exceed $L_{\rm 3-20 keV}\la3\cdot10^{33}(d/1kpc)^2$ ergs/s (where $d$ is
the distance to the black hole). Near the maximum of the background star
amplification by the microlens (July 6, 1999), the upper limit on the X-ray
flux corresponds to an X-ray luminosity $L_{\rm 3-20
  keV}\la7\cdot10^{33}(d/1kpc)^2$ ergs/s.  
\keywords{RXTE/PCA, black holes, 
   X-ray binaries, gravitational lensing, interstellar medium}
}
\maketitle

\section*{INTRODUCTION}

\indent

The possibility of gravitational microlensing by stars in the Galaxy --
an apparent increase in the optical brightness of the star, caused by the
gravitational lensing by an object crossing the line of sight -- has been
discussed since 1970-1980th (see Paczynski 1996 for a review).
However, because of the great technical problems in observing such
events, the significant progress in this field was achieved only recently
 (MACHO and OGLE groups). So far, more than a thousand 
microlensing events have been discovered (see, e.g. Alcock et al., 2000,
 Wozniak et al., 2001).

During the crossing of the line of sight by a lensing object the brightness
of the background star is changing in a specific way (see, e.g. 
Paczynski, 1986). The timescale of a microlensing event depends on the lens
mass, the distances to the lens and to the background star, and 
the transverse velocity of the lens. Observational studies of the
probability distribution of microlensing timescales show that the
most frequent events occur on time scales of the order of 100 days, which
indicates that the typical lens mass is of the order of a solar mass (see
Paczynski, 1996; Alcock et al., 2000).

Recently, a more detailed investigation of the longest observed microlensing
event \ogle (Mao et al., 2001) led to the conclusion that the
lensing object was a black hole candidate. It was demonstrated that the mass of the lens
$M\sim4.4M_{\odot}$ if the lens distance is $d\sim$6 kpc, and  
$M\sim200M_{\odot}$ if the lens distance is $\sim$500 pc (assuming that
the background star resides in the bulge, $d_{star}\sim7$kpc). In any case, 
the lens mass appears to be beyond the mass limit for neutron stars.
Therefore, the observations of the microlensing event \ogle strongly indicate
the presence of a black hole in the direction of: $l=2.46, b=-3.505$, or
$\alpha=18^h05^m05.35^s,  \delta=-28^\circ34\arcmin42.5\arcsec$. The maximum
amplification of the background star brightness occurred on July 6, 1999
($\sim$ TJD~11365).

In this Letter we derive an upper limit on the X-ray luminosity of the
black hole \ogle using the data of RXTE/PCA scans over the Galactic 
Center region in 1999-2000.

\begin{figure*}
\pspicture(0.,15)(0.,21)
\rput(8,15){\epsfxsize 16cm \epsffile[0 170 550 600]{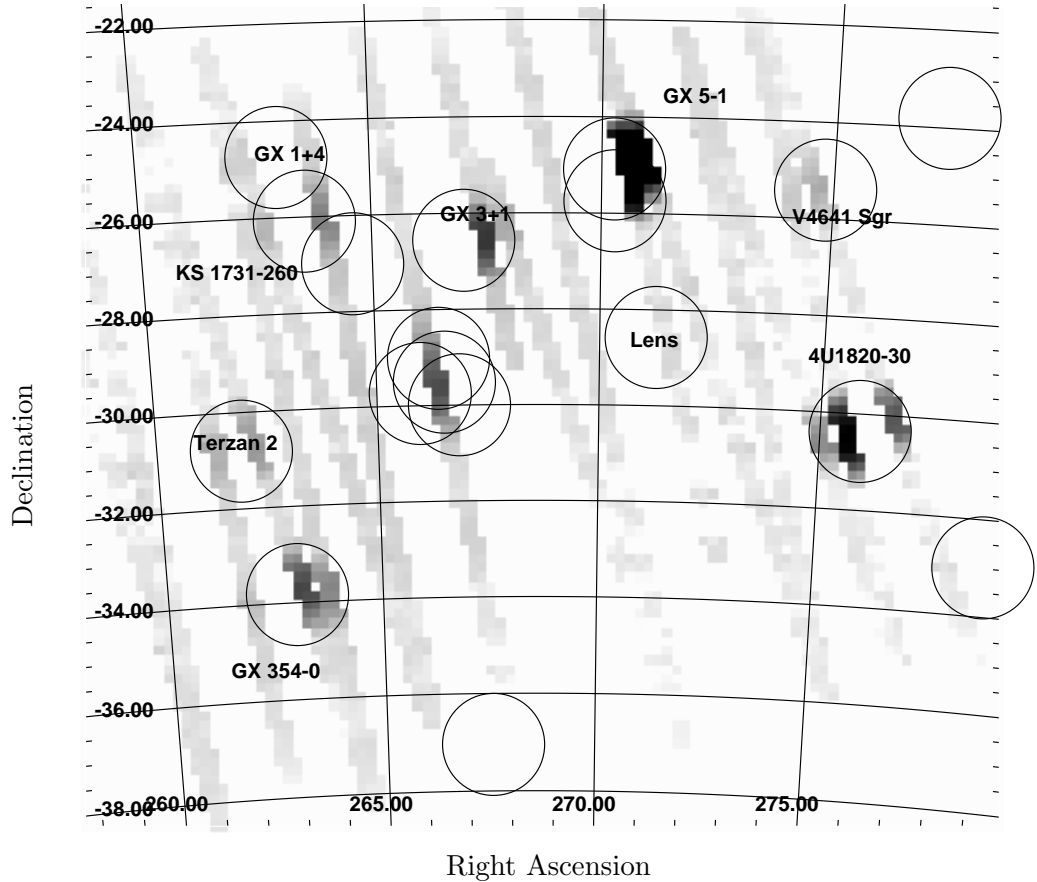}}
\rput(9,9.5){\sf\large Right Ascension}
\rput(2.0,15.0){\rotateleft{\sf\large Declination}}
\endpspicture
\vspace{6cm}
\caption{Map of the Galactic Center region according to the PCA scan 
observations on July 7, 1999. Solid circles represent the regions of the 
bright sources ``illumination'' (see text).\label{gc_map}}
\end{figure*}

\section*{OBSERVATIONS AND RESULTS}

\indent

There are three scientific instruments aboard the Rossi X-ray Timing Explorer
(RXTE) observatory: two coaligned spectrometers PCA (3--60 keV) and HEXTE 
(20-200 keV) with fields of view 1$^\circ$, designed for detailed studies of
X-ray sources, and an All Sky Monitor --  ASM (1-12 keV), which allows one to
follow their long term behavior. In Feb. 1999, a campaign of scans over the
Galactic Center region with the PCA spectrometer was initiated. This
instrument has a very large effective area ($\sim$6500 cm$^2$), and 
the uncertainty in measuring the flux for sources detected in a single scan is
of the order of 1--2 mCrabs. Therefore, it appeared that the data of PCA
scan observations effectively complement the
ASM monitoring results. Note that even in the scanning mode, PCA provides a
sensitivity that is higher by an order of magnitude than that of ASM. This
advantage in sensitivity becomes of particular importance if a given event
is short so that ASM cannot provide us good statistics. 

\begin{figure}
\epsfxsize=8cm
\epsffile[20 150 570 700]{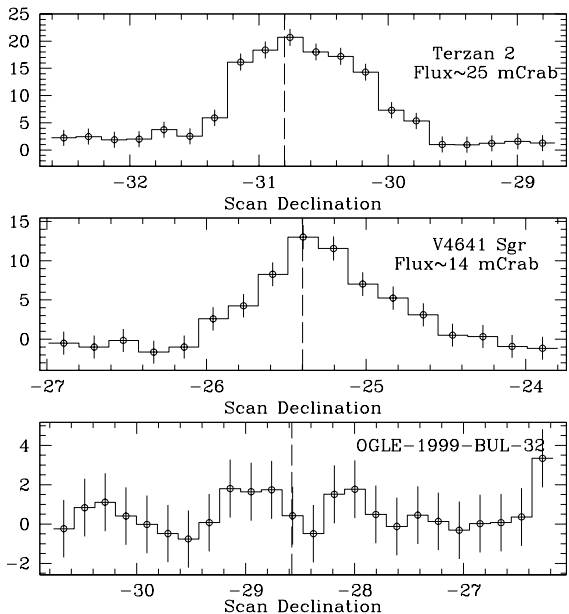}
\caption{Slices of the PCA scans over Terzan 2, V4641 Sgr and the 
microlens \ogle 7 ÌÀÌÑ 1999 Ç.\label{slices}}
\end{figure}

We analyzed approximately a hundred scanning PCA observations covering the
period from Feb. 1999 to Mar. 2000. One of these scans was performed on July
7, 1999,  i.e. very close to the maximum of the lensing amplification of the
background star by \ogle.

The PCA data were analyzed with the help of standard tools of the LHEASOFT
package. For the background estimation we used model $L7/240$.

Fig. \ref{gc_map} presents a map of the Galactic Center region, obtained
from the PCA scan observations performed on July 7, 1999 (the effective
energy band 3--20 keV). In constructing the map, the flux collected by the
PCA over a 1 sec time interval was assigned to the celestial position
corresponding to the center of the PCA field of view. This method of map
construction causes 1$^\circ$ regions around bright objects to appear
``illuminated'' in accordance to the collimator radial response function. In
Fig. \ref{gc_map} one can clearly see such ``illumination'' regions (they
are indicated by solid circles) encircling well known bright objects GX5-1,
4U1820-30, GX3+1, GX354-0. Also apparent are comparably weak sources,
including Terzan 2 and the accreting black hole in the high-mass binary system
V4641 Sgr (the 3--20 keV X-ray flux from these sources was $\sim$25 mCrab and 
$\sim$14 mCrab correspondingly). In Fig. \ref{slices} we present the slices
of PCA spectrometer scans over the positions of Terzan 2, V4641 Sgr and the 
microlens \ogle. It is seen that the flux from the position of \ogle in this
observation did not exceed 1-2 mCrab.

\begin{figure*}
\pspicture(0.,15)(0.,21)
\rput(8,15){\epsfxsize 16cm \epsffile[0 240 550 600]{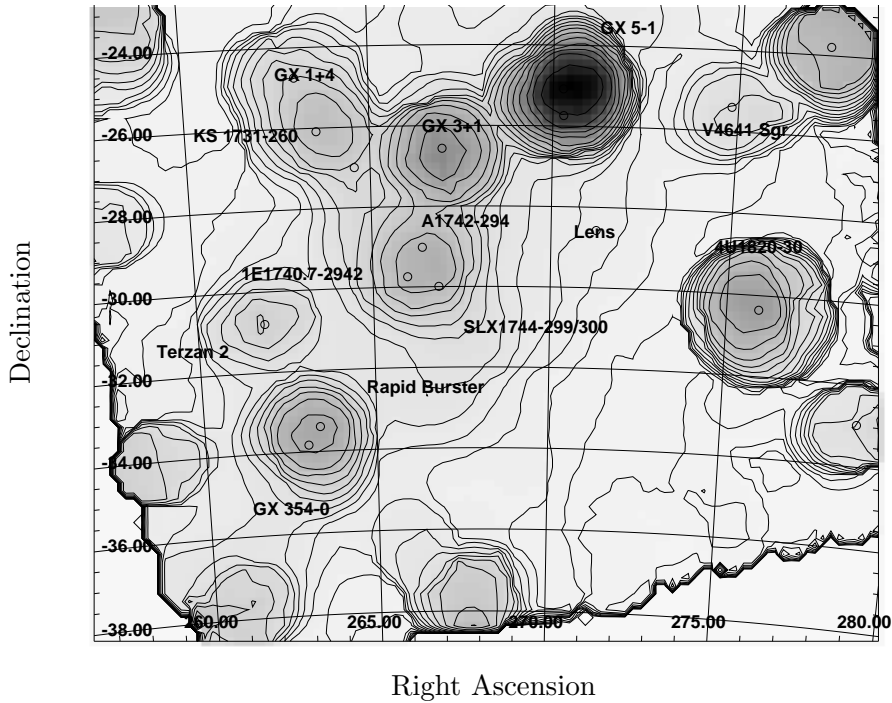}}
\rput(9,10){\sf\large Right Ascension}
\rput(2.7,15.0){\rotateleft{\sf\large Declination}}
\endpspicture
\vspace{6cm}
\caption{Map of the Galactic Center region by PCA scan observations, averaged
over Feb.1999-Mar.2000.\label{gc_map_all}} 
\end{figure*}

In Fig. \ref{gc_map_all} we present a map of the Galactic Center region
obtained using all available PCA scan observations during the period
Feb.1999-Mar.2000. It should be noted here that despite the dramatic
improvement in statistics (about 100 PCA scan observations were co-added),
it proves impossible to improve significantly the upper limit on the source
flux, because of the systematic uncertainties in the background subtraction
and the influence of the Galactic diffuse emission. The upper limit on the
X-ray flux from \ogle in this case $F_x\la$1 mCrab.

\section*{DISCUSSION}
In the previous section we derived an upper limit on the 3--20 keV 
X-ray flux from \ogle of the order of $\la$1 mCrab (Feb.1999-Mar.2000),
which corresponds to an X-ray luminosity $L_{\rm 3-20
  keV}\la3\cdot10^{33}$(d/1kpc)$^2$ ergs/s, where $d$ is the distance to the
black hole. 

This upper limit can be important for models considering the
binary system harboring the black hole \ogle.

Observations of X-ray Novae -- low mass binaries with black holes -- have
shown that there are thousands of recurrent transient X-ray sources in the
Galaxy. These typically demonstrate X-ray activity only during several
months every 50-70 years (see, e.g. Tanaka, Shibazaki, 1996, for a
review). The discovery of an X-ray outburst of V4641 Sgr revealed that black
holes, which are unobservable in X-rays at particular times, can also exist
in high mass binary systems (Orosz et al. 2001). However, it is clear that
a bright optical companion of the black hole \ogle would have been detected
during the optical microlens observations. Therefore, the hypothesis of a high
mass binary harboring the black hole \ogle appears very unlikely. However,
the question of whether this black hole is in a low mass binary 
system still remains open given the possibility that the distance of the
source is large ($d\ga$3--4 kpc). In the case of a near ($d\sim$500 
pc), massive ($M\sim200M_{\odot}$) black hole, the microlensing observations
would have revealed the optical companion if it were more massive than
$\ga0.2M_{\odot}$. Therefore, we are coming to the conclusion that such 
a massive black hole should be single.

Observations have revealed a small amount of molecular hydrogen,
$N_HL\sim5\cdot10^{18}$cm$^{-2}$, in the direction of
the black hole \ogle (Dame et al., 1987). However, a much larger amount of
neutral hydrogen is present in this direction  -- $N_HL\sim3\cdot10^{21}$
cm$^{-2}$ (Dickey, Lockman, 1990). This means that at energies lower than
0.5-1 keV the possible X-ray or UV source would be strongly absorbed.

According to McKee, Ostriker (1977), most of the volume of the
Galaxy disk is filled by a rarefied gas with embedded molecular clouds and
clouds of neutral hydrogen. The relatively low column density of molecular
hydrogen in the direction of the source indicates that the probability of the
black hole residing in a molecular cloud is rather low. However, the black hole
might be in a neutral hydrogen cloud. In this case it could accrete matter
from the interstellar medium at a rate sufficient for it to be detected in
EUV or in X-rays. 
 
Turbulent velocities and rotation of the gas in an interstellar cloud
should lead to the formation of a disk during the accretion onto the
compact object. If the innermost regions of the accretion flow are not
 dominated by advection (not an ADAF-like flow), then the emission of
the accretion disk can be described by the multicolor disk model 
(Shakura, Sunyaev, 1973). If the accretion rate is small, which is presumably
true in our case, the maximal temperature of the accretion disk does not
exceed 0.5 keV, hence most of the energy is radiated outside
our energy band (3--20 keV). This makes it very difficult to place an upper
limit on the bolometric luminosity of the black hole, using our measurements
 in 3--20 keV energy band. A more accurate
estimate could be made if CHANDRA of XMM-NEWTON data were available
(the effective energy band is $\sim$0.1-10 keV). 

It was shown in the previous section that one of the scanning PCA observations
was performed 1 day after the maximum of the background star brightness
amplification took place (cf. with the characteristic time of the microlensing
event $\sim$640 days). The upper limit on the X-ray flux from the
lens/background star during this observation is $F_x\la$2.5 mCrab
(2$\sigma$). Taking into account that the background star amplification was
then $\sim$12, we can put an upper limit on the X-ray luminosity of this
background star (assuming that it is located in the Galactic bulge with
$d\sim7$kpc): $L_x\la2\cdot10^{34}$ ergs/s. This upper limit is not very
stringent, as it exceeds the luminosity of the most powerful X-ray flares
of the Sun.

The fact of detection of a massive object that is very faint in optics and 
X-rays tells us that black holes in the interstellar medium are not
exotic phenomena. It is obvious that some of them can from time to time pass
through a dense molecular clouds and clouds of interstellar gas. When such
an event takes place, these objects can start emit X-rays and can be
detectable at the level of sensitivity of the RXTE observatory.

Unfortunately, a source with a luminosity $L_x>10^{37}$ ergs/s can
heat the surrounding medium to high temperatures in quite a short 
period of time, and that will cause an outflow of matter. This, in turn,
will lead to a turn-off of the accretion even if the original density 
of the cloud was high enough (Sunyaev, 1978).

\begin{acknowledgements}
Authors thank Marat Gilfanov, Conrad Cramphorn, Shude Mao and Sergei
Sazonov for usefull 
discussions. This research has made use of data obtained through the High
Energy Astrophysics Science Archive Research Center Online Service, provided 
by the NASA/Goddard Space Flight Center.

\end{acknowledgements}

\section*{óðéóïë ìéôåòáôõòù}
\indent

{\it Alcock C.}//
Astroph.J. 2000, 541, 734

{\it Dickey J., Lockman F.}//
Ann.Rev.Ast.Astr. 1990, 28, 215

{\it Dame T., Ungerechts H., Cohen R. et al.}//
Astroph.J. 1987, 322, 706

{\it  McKee C., Ostriker J.}//
Astroph.J. 1977, 218, 148

{\it Mao S.}//
MNRAS 2001, submitted (astro-ph/0108312)

{\it Orosz J., Kuulkers E., van der Klis M.}//
Astroph.J. 2001, 555, 489

{\it Paczynski B.}//
Astroph.J. 1986, 304, 1

{\it Paczynski B.}//
Astroph.J. 1991, 371, 63L

{\it Paczynski B.}//
Ann.Rev.Ast.Astr. 1996, 34, 419

{\it Sunyaev R.}//
SvAstron.Lett. 1978, 4, 75

{\it Tanaka Y., Shibazaki N.}//
Ann.Rev.Ast.Astr. 1996, 34, 607

{\it Shakura N., Sunyaev R.}//
Astron.Astroph. 1973, 24, 337

{\it Wozniak P. et al.}//Acta Astron., submitted (astro-ph/0106474)

\end{document}